\documentclass[12pt]{iopart}
\usepackage{graphicx}
\begin{document}
\title{Pairwise Quantum Correlations for Superpositions of Dicke States}

\author{Zhengjun Xi$^{1,2}$, Heng-Na Xiong$^2$, Yongming Li$^1$, and Xiaoguang Wang$^2$}

\address{$^1$ College of Computer Science, Shaanxi
Normal University, Xi'an, 710062, P.R. China}
\address{$^2$ Zhejiang Institute of
Modern Physics, Department of Physics, Zhejiang University, HangZhou
310027, P. R. China.} \eads{\mailto{snnuxzj@gmail.com},
\mailto{xgwang1208@zju.edu.cn}}

\begin{abstract}

Pairwise correlation is really an important property
for multi-qubit states. For the two-qubit X states
extracted from Dicke states and their superposition states, we obtain
a compact expression of the quantum discord by numerical check. We then apply
the expression to discuss the quantum correlation of the reduced
two-qubit states of Dicke states and their superpositions, and the
results are compared with those obtained by entanglement of
formation, which is a quantum entanglement measure.
\end{abstract}

\begin{figure}[ptb]
\begin{center}
\includegraphics[
height=4cm]{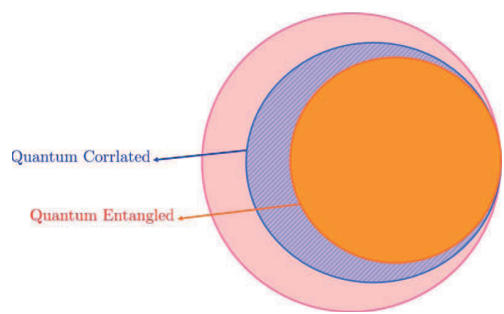}
\end{center}
\caption{(Color online) Schematic configuration of the relations
between quantum correlation and quantum entanglement. The pink
circle represents the whole set of quantum states. They can be split
into quantum correlated ones (the blue circle) and classical
correlated ones. They can also be divided into quantum entangled
ones (the red circle) and quantum separable ones. In general, the
states with quantum entanglement must be quantum correlated.
However, a quantum correlated state may not be quantum entangled.
That is, there exists a class of quantum separable states which are
also quantum correlated, as shown by the shadow area. }
\label{Fig-relation}
\end{figure}

\section{Introduction}

\ \ \ \ \ It is well known that quantum entanglement is a typical
quantum computation resource and plays an important role in quantum
information processing\thinspace ~\cite{Nielsen00,Vedralrmp02}. Much
attention has been paid to detect and measure the quantum
entanglement (see \thinspace\cite{Plenio07,Horodecki09,Wootters98}
and references therein). Traditionally, it is believed that quantum
entanglement is synonymous with quantum correlation. However, it is
not always the case. Many works showed that there exists a more
general quantum
correlation\thinspace\cite{Groisman05,Horodecki05,Bennett99,Lanyon08}
besides entanglement, as shown in figure \ref{Fig-relation}. That
is, some separable states can also possess quantum correlation,
which can improve some computational tasks \cite{Datta05}. Such the
general quantum correlation is quantitatively characterized by
quantum discord (QD)\thinspace
\cite{Ollivier02,Zurek00,Vedral02,Vedral03}, which is defined from
the quantum measurement perspective. Using this quantity, it was
proved that almost all quantum states actually have quantum
correlations~\cite{Ferraro}. Most recently, operational interpretations of quantum discord were proposed, where
quantum discord was shown to be a quantitative measure about the performance in quantum
state merging~\cite{Cavalcanti,Madhok}.
Up to now, QD has been widely
studied in many fields\cite{Werlang09,Maziero10,Fanchini09,Piani09,Dill08,Sarandy09,chen10,
Werlang0911,Alireza09,Luo08,Maziero1002,Datta09,Wu09,Csar08,
Mazhar10,Brodutch10,Saitoh08,Kaszlikowski,Modi2010,Fanchini2460,Chen,Hao,Xu,Xupra,Auccaise}.

Calculating QD involves an optimization process over all possible
quantum measurements. So far, a general analytical expression of QD
still lacks even for the two-qubit states. Some analytical results
were shown only for a special subset of two-qubit X-states\thinspace
\cite{Ollivier02,Vedral02,Fanchini09,Luo08,Maziero1002,Mazhar10}.
In this article, we consider
a kind of two-qubit X-states with exchange and parity symmetries,
whose off-diagonal elements are complex, as shown in
equation~(\ref{X structured}). We give an upper bound for the
expression of QD. We find that the expression does not depend on the
arguments of the complex off-diagonal elements.

A typical example of the two-qubit X states shown in
equation~(\ref{X structured}) is the reduced two-qubit states of
Dicke states and their superposition states. Two motivations inspire
us to study the pairwise (or two-qubit) quantum correlation
properties of such kinds of states. One comes from the experimental
perspective. Dicke states are fundamental multiqubit states. They
can be realized both in atomic systems~\cite{Molmer99,Lemr09} and in
photonic systems~(\cite{Mattews09} and references therein). Many
multi-qubit states are based on them, such as W
states~\cite{Zeilinger97,Dur00}, GHZ states~\cite{Greenberger95} and
spin coherent states (SCSs)~\cite{Radcliffe71,Kitagawa93}, etc. The
importance of the Dicke states in experiment has attracted much
theoretical studies~\cite{Agrawal06,Toth07}. Here we would like to
examine their pairwise quantum correlation properties in terms of
QD. The other motivation comes from the theoretical aspect. The
exchange symmetry of a Dicke state ensures that its two-qubit
reduced state can be extracted randomly from the global state. That
is, the reduced state of the $i$-th and $j$-th qubits
$\rho_{ij}={\rm{Tr}}_{\overline{{ij}}}(\rho_{\rm{global}})$ is
invariant for arbitrary positions of $i$ and $j$. Once the two-qubit
reduced state $\rho_{{ij}}$ is quantum correlated, all of its
possible two-qubit states are quantum correlated. Therefore, to some
extent, the existence of the two-qubit quantum correlation
sufficiently reflects the multi-qubit quantum correlation.
Substantial efforts have been made to quantify
multi-qubit quantum correlation in terms of its pairwise correlations
\cite{Dill08,Sarandy09,Chen,Hao,Linden02,Walck08,Diosi04,Wang02}. All theses
works show that pairwise correlation is really an important property
for multi-qubit states.

This paper is organized as follows. In Sec.\thinspace\ref{sec:qc},
we give the basic concepts of QD and EoF. The latter is a quantum
entanglement measure. In Sec.\thinspace\ref{sec:Xstate}, for the
two-qubit X states with exchange and parity symmetries, whose
off-diagonal elements are complex, an upper bound of QD is
analytically derived. In Sec.\thinspace\ref{sec:symmetric}, we make
a comparison between the pairwise QD (PQD) and pairwise EoF (PEoF)
for Dicke states and their superposition states.

\section{Quantum Correlations\label{sec:qc}}
\subsection{Quantum discord}
In classical information, the mutual information measures the correlation
between two random variables $A$ and $B$
\begin{equation}
\mathcal{I}_c(A:B)=H(A)+H(B)-H(AB),\label{eq:ccM1}
\end{equation}
where $H(X)=-\sum_xp_x\log_2p_x$ are the Shannon entropies for the variable $X (X=A,B)$ and $H(A,B)=-\sum_{a,b}p_{a,b}\log_2p_{a,b}$
is the joint system $AB$, $p_{a,b}$ is the joint probability of the variables $A$ and $B$ assuming the values
$a$ and $b$, respectively, and $p_a = \sum_b p_{a,b} (p_b =\sum_a p_{a,b})$ is the marginal probability of the variable $A (B)$ assuming the value $a (b)$~\cite{Cover}.
The classical mutual information can also be expressed in terms of the conditional entropy as
\begin{equation}
\mathcal{J}_c(A:B)=H(A)-H(A|B),\label{eq:ccM2}
\end{equation}
where $H(A|B) =\sum_{a,b} p_{a,b} \log_2 p_{a|b}$ is the conditional entropy of the variable $A$ given that variable $B$ is known, and $ p_{a|b}=p_{a,b}/p_b$ is the conditional probability.

In quantum information, to obtain a quantum version of Eq.(\ref{eq:ccM2}),
Ollivier and Zurek employed a complete set of perfect
orthogonal projective measurements $\{\Pi_{k}^{B}\}$ with
$\sum_{k}\Pi_{k}^{B}=I_B$ on the subsystem $B$. The reduced state of subsystem $B$ after the measurement is given as
$\rho_{A|\Pi_{k}^{B}}= \frac{1}{p_{k}}\mathrm{Tr}_B\left[(I^{A}\otimes\Pi_{k}^{B})
\rho^{AB}( I^{A}\otimes\Pi_{k}^{B})\right] $, where $p_{k}=\mathrm{Tr}_{AB}(I_{A}\otimes\Pi_{k}^{B}\rho^{AB})$ is probability
for the measurement of the $k$th state in the subsystem $B$. Then, for a known subsystem $B$, one can define conditional entropy of subsystem $A$, $S_{\{\Pi_{k}^{B}\}}(\rho_{A|B}):=\sum_{k}p_{k}S(\rho_{A|\Pi_{k}^{B}})$. Then, for a bipartite quantum state $\rho^{AB}$, the quantum analogues for Eqs.(\ref{eq:ccM1}) and (\ref{eq:ccM2}) are given as
\begin{equation}
\mathcal{I}(\rho^{AB})=S(\rho^{A})+S(\rho^{B})-S(\rho^{AB}),\label{mutual infromation}%
\end{equation}
and
\begin{equation}
\mathcal{J}(\rho^{AB})=S(\rho^{A})-S_{\{\Pi_{k}^{B}\}}(\rho_{A|B}).\label{eq:qqM2}
\end{equation}
where $\rho^{A(B)}=$Tr$_{B(A)}\rho^{AB}$ is the reduced density
matrix for $A$ ($B$), and
$S(\rho^{X})=-$Tr$(\rho^{X}\log_{2}\rho^{X})$ ($X=A,~B$) is the von
Neumann entropy~\cite{Nielsen00}.
In general, Eqs.(\ref{mutual infromation}) and (\ref{eq:qqM2}) are not equivalent~\cite{Ollivier02,Zurek00}, their difference is defined as quantum discord (QD), which is a more general characterization of quantum correlation, namely,
\begin{equation}
\mathcal{D}(\rho^{AB}):=\mathcal{I}(\rho^{AB})-\max_{\{\Pi_{k}^{B}\}}\mathcal{J}(\rho^{AB}).\label{quantum discord}
\end{equation}

Quantum mutual information (\ref{mutual infromation}) is usually used to quantify
the total amount of the correlation between the two subsystems $A$
and $B$~\cite{Groisman05,Vedral02}. In particular, if one considers positive operator valued measure (POVM) on subsystem $B$ and maximize the equation~(\ref{eq:qqM2}),
it may be viewed as classical correlation which is defined by Henderson and Vedral~\cite{Vedral02}.
Hamieh $etal$ shown that the projective measurement is POVM that maximizes~(\ref{eq:qqM2}) for two-qubit system.
In our article, we will only compute quantum discord for two-qubits states in terms the original definition~\cite{Ollivier02,Zurek00}.

An important property of QD is that, even if $\rho^{AB}$ is a
separable state, its QD may be nonzero. That is, QD captures more
general quantum correlation than entanglement, as shown in
figure~\ref{Fig-relation}. It has been experimentally proved that
quantum separable states with nonzero QD have played an important
role in quantum computation\thinspace \cite{Lanyon08,Datta05}.

\subsection{Entanglement of formation}

For an arbitrary two-qubit state $\rho^{AB}$, concurrence is one of
the most widely used measurements of entanglement, which is defined
as\thinspace \cite{Wootters98},
\begin{equation}
C(\rho^{AB})=\max\Big\{0,\sqrt{\lambda_{1}}-\sqrt{\lambda_{2}}-\sqrt{\lambda_{3}}-\sqrt{\lambda_{4}}\Big\},
\end{equation}
where $\lambda_{1},~\lambda_{2},~\lambda_{3},~\lambda_{4}$ are the
decreasing ordered eigenvalues of the matrix $R=\rho^{AB}\left(
\sigma_{y}\otimes \sigma_{y}\right)  (\rho^{AB})^{\ast}\left(
\sigma_{y}\otimes\sigma _{y}\right)  $ with $\sigma_{y}$ the pauli
matrix, and $(\rho^{AB})^{\ast}$ is the complex conjugation of
$\rho^{AB}$. To compare quantum entanglement with QD, in the
following, we would like to use the entropy function of concurrence
(EoF) as the entanglement measure. The EoF is defined
as\thinspace\cite{Wootters98}
\begin{equation}
E_{F}(\rho^{AB})=\min_{\{p_{i},|\phi_{i}\rangle
\}}\Big[\sum_{i}p_{i}S(\mathrm{Tr}_A(|\phi_{i}\rangle\langle\phi_{i}|)\Big],\label{formation meas}
\end{equation}
where the minimum is taken over all possible decompositions ${\{p_{i},|\phi_{i}\rangle\}}$
with $\rho ^{AB}=\sum_{i}p_{i}\left\vert
\phi_{i}\right\rangle \left\langle \phi _{i}\right\vert $.

For a general two-qubit state, the EoF can be expressed by
concurrence $C$\thinspace\cite{Wootters98}%
\begin{equation}
E_{F}(\rho^{AB})=H\left(  \frac{1+\sqrt{1-C^{2}}}{2}\right)  ,
\label{formation}%
\end{equation}
where the binary entropy is
\begin{equation}
H(x)=h(x)+h(1-x), \label{binary_en}%
\end{equation}
with the function $h(x) =-x\log_{2}x (0 \leq x\leq 1).$

There is a closed relation between QD and EoF for a
bipartite pure state. Since the conditional density operators
$\rho_{A|\Pi_{k}^{B}}$ is also a pure state, then the quantum
conditional entropy $S_{\{\Pi_{k}^{B}\}}(\rho_{A|B})=0$. Therefore,
QD is reduced to EoF, which is equal to the von Neumann entropy of
the subsystem $A$, i.e.,
\begin{equation}
\mathcal{D}(\rho^{AB})=E_{F}(\rho^{AB})=S\left(  \rho^{A}\right)  .
\end{equation}
However, for a general mixed state, the two concepts show totally
different behaviors due to the difference between their definitions.
The QD describes the quantum correlation in terms of quantum
measurements. It is based on the idea that a classical correlated
state remains unchanged under a quantum measurement on one of the
subsystems. However, the quantum entanglement (such as EoF) is
defined mathematically opposite to quantum separability.

\section{\textbf{Quantum discord for two-qubit X states with exchange and parity symmetries \label{sec:Xstate}}}

Here, we would like to consider the case that the
off-diagonal elements of the density matrix are complex as shown in
equation~(\ref{X structured}). The density matrix of a two-qubit
state with exchange and parity symmetries takes the
form~\cite{Wang02}
\begin{equation}
\rho^{AB}=\left(
\begin{array}
[c]{cccc}%
v_{+} & 0 & 0 & u^{\ast}\\
0 & y & y & 0\\
0 & y & y & 0\\
u & 0 & 0 & v_{-}%
\end{array}
\right)\label{X structured}%
\end{equation}
in the basis $\{\left\vert 00\right\rangle ,\left\vert
01\right\rangle,\left\vert 10\right\rangle ,\left\vert
11\right\rangle \}$, where the elements $v_{+}$, $v_{-}$ and $y$ are
real numbers, $u$ is a complex number, and $u^{\ast}$ is the complex
conjugate of $u$.

First, the joint entropy is easily given by

\begin{equation}
S(\rho^{AB})=h(\lambda_{0})+h(\lambda_{+})+h(\lambda_{-}), \label{S_AB (X)}%
\end{equation}
with $\lambda_{i}^{'}$s the eigenvalues of $\rho^{AB}$
\begin{equation}
\lambda_{0}=2y,~~
\lambda_{\pm} =\frac{1}{2}\left(  v_{+}+v_{-}\pm\sqrt{(v_{+}-v_{-}%
)^{2}+4|u|^{2}}\right)  .
\end{equation}
Meanwhile, after tracing over the degree of the qubit $A$, we obtain
the reduced density matrix for the qubit $B$
\begin{equation}
\rho^{B}=\Tr_{A}(\rho^{AB})=\left(
\begin{array}
[c]{cc}%
v_{+}+y & 0\\
0 & v_{-}+y
\end{array}
\right)  , \label{Marginal_DMAB}%
\end{equation}
with the von Neumann entropy%
\begin{equation}
S(\rho^{B})=H\left(  v_{+}+y\right)  . \label{S_B (X)}%
\end{equation}

The quantum conditional entropy $S_{\{\Pi_{k}^{B}\}}(\rho_{A|B})$
involves all possible one qubit projective measurements
$\{\Pi_{k}^{B}\}$. Here we choose the measurements
\begin{equation}
\Pi_{\pm}^{B}=\frac{1}{2}\left(  I\pm\vec{n}\cdot\vec{\sigma}\right)
,
\end{equation}
where $\vec{n}=(\sin\theta\cos\phi,\sin\theta\sin\phi,\cos\theta)$
with $\theta\in\left[  0,\pi\right]  $ and $\phi\in\lbrack0,2\pi)$.
Then the conditional density operators
$\rho_{\pm}^{A}\equiv\rho_{A|\Pi_{\pm}^{B}}$ are
\begin{equation}
\fl \rho_{\pm}^{A}\left(  \theta,\phi\right)
=\frac{1}{2p_{\pm}\left( \theta\right)  }
  \left(
\begin{array}
[c]{cc}%
\left(  v_{+}+y\right)  \pm\left(  v_{+}-y\right)  \cos\theta & \pm
(ye^{-i\phi}+u^{\ast}e^{i\phi})\sin\theta\\
\pm(ye^{i\phi}+ue^{-i\phi})\sin\theta & \left(  v_{-}+y\right)
\mp\left( v_{-}-y\right)  \cos\theta
\end{array}
\right)
\end{equation}
with the probabilities
\begin{equation}
p_{\pm}\left(  \theta\right)  =\frac{1}{2}\left[
1\pm(v_{+}-v_{-})\cos \theta\right]  .
\end{equation}
Obviously, $\rho_{-}^{A}\left(  \theta,\phi\right)
=\rho_{+}^{A}\left( \pi-\theta,\phi\right)  $, and $p_{-}\left(
\theta\right)  =p_{+}\left(
\pi-\theta\right)  $. Therefore, the conditional entropy is given by%
\begin{eqnarray}
S_{A|B}\left(  \theta,\phi\right) & \equiv &
S_{\{\Pi_{k}^{B}\}}(\rho
_{A|B}\left(  \theta,\phi\right)  )\nonumber\\
  &=& p_{+}(\theta)H\left(  {\large
\frac{1+\kappa(\theta,\phi)}{2}}\right)
{\large +p_{-}(\theta)H}\left({\large \frac{1+\kappa(\pi-\theta,\phi)}{2}%
}\right){\large,}
\end{eqnarray}
where $\kappa(\theta,\phi)  $ is one of the eigenvalues
of conditional state $\rho_{+}^{A}(\theta,\phi)$, and%
\begin{eqnarray}
\fl \kappa^{2}(\theta,\phi) =\frac{1}{p_{+}^{2}(\theta)}\Big[\frac{1}{4}((v_{+}-v_{-})
+(1-4y)  \cos\theta)^{2}+(|u|^{2}+y^{2}+2y{Re}(u)e^{-i2\phi})
\sin^{2}\theta)\Big].\nonumber
\end{eqnarray}

The following work is to minimize the conditional entropy
$S_{A|B}(\theta,\phi)$ over the parameters of $\theta$
and $\phi$. First, we note that the probabilities
$p_\pm(\theta)$ are independent of $\phi$, taking the
derivative $S_{A|B}(\theta,\phi) $ over $\phi$. From the
equation $\frac{\partial S_{A|B}(\theta,\phi)
}{\partial\phi}=0$, we get the value of $\phi$ making
$S_{A|B}(\theta,\phi)$ minimum at
\begin{equation}
\phi_m=\frac{1}{2}\arg (u)\nonumber\ {\rm{or}}\ \ \frac{1}{2}\arg(u)+\pi,\label{min_phi}%
\end{equation}
where $\arg(u)  \in\lbrack0,2\pi)$ is the argument of
the complex number $u$. Thus,
\begin{equation}
\min_{\phi}S_{A|B}(\theta,\phi) =p_{+}(\theta)H(
\frac{1+\widetilde{\kappa}(\theta)}{2})+p_{-}(\theta)H(\frac{1+\widetilde{\kappa}(\pi-\theta)}{2}),\label{QCE}
\end{equation}
with
\begin{equation}
\fl\widetilde{\kappa}^{2}(\theta)=\frac{1}{p_{+}^{2}(\theta)}\Big[\frac{1}{4}((v_{+}-v_{-}) +(1-4y)  \cos\theta)^{2}
  +(|u|^{2}+y^{2}+2y\vert u\vert)\sin^{2}\theta)\Big],\label{k(theta)}%
\end{equation}
which is independent of the argument of $u$. So far, the
minimization of $S_{A|B}(\theta,\phi)  $ over $\phi$ is
done completely. However, its optimization over $\theta$ is so
difficult that we only give an upper bound
\begin{equation}
\min_{\theta,\phi}S_{A|B}(\theta,\phi) \leq\min\{S_{0},S_{1}\}  ,\label{upper}%
\end{equation}
where
\begin{equation}
S_{0}=S_{A|B}(\theta=0,\phi_m),~~~S_{1}=S_{A|B}(\theta=\frac{\pi}{2},\phi_m) .
\end{equation}
They are obtained at $\theta=0,\frac{\pi}{2}$ in terms of the fact
that $S_{A|B}\left(  \theta,\phi_m\right) =S_{A|B}\left(
\pi-\theta,\phi_m\right)  $. That is, $S_{A|B}\left( \theta
,\phi_m\right)  $ is symmetric around $\theta=\frac{\pi}{2}$. Thus
$S_{0}$ and $S_{1}$ are two (not the only two) extremes of the
conditional entropy $S_{A|B}\left( \theta,\phi\right)$ over the
parameters $\theta$ and $\phi$. From equations\thinspace(\ref{QCE})
and\thinspace (\ref{k(theta)}), the explicit expressions for $S_{0}$
and $S_{1}$ is easily
derived as%
\begin{equation}
\fl S_{0}=(v_{+}+y)H\left(  \frac{1+\kappa_{+}}{2}\right)
+(v_{-}+y)H\left( \frac{1+\kappa_{-}}{2}\right),~~
S_{1}=H\left(  \frac{1+\kappa_{1}}{2}\right),\label{S_0 and S_1}%
\end{equation}
with%
\begin{equation}
\kappa_{\pm}=\frac{|v_{\pm}-y|}{v_{\pm}+y},\ \ \kappa_{1}=\sqrt
{(v_{+}-v_{-})^{2}+4(y+|u|)^{2}}.\label{k_pm,k_1}%
\end{equation}

\begin{figure}[ptb]
\begin{center}
\includegraphics[
height=5cm, width=8cm ]{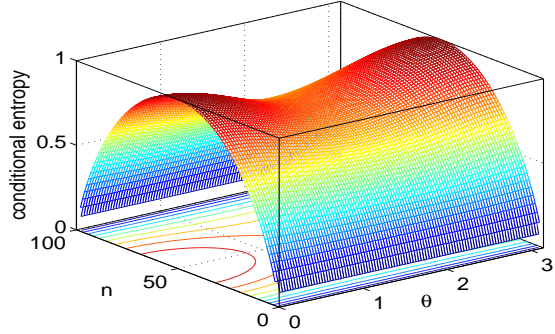}
\end{center}
\caption{(Color online) Conditional entropy
$\min_{\phi}S_{A|B}(\theta,\phi)$ [as shown in equation~(\ref{QCE})]
of Dicke state $|n\rangle_{N}$ as a function of $n$ and $\theta$ for
fixed $N=100$. The minimum of $\min_{\phi}S_{A|B}(\theta,\phi)$ for
fixed $n$ and $N$ is obtained at $\theta=\pi/2$.}
\label{Fig_Dicke_3D}
\end{figure}

Many works have shown that the upper bound (\ref{upper}) is tight,
i.e., $\min_{\theta,\phi}S_{A|B}\left( \theta,\phi\right)
=\min\left\{ S_{0},S_{1}\right\}$ \cite{Fanchini09,Luo08,Mazhar10}.
However, for some two-qubit X-state, this upper bound is not
necessary reachable, a counterexample is given in \cite{XMLuI1009}
(see the equation (18) therein). Fortunately, for the two-qubit
density matricies which have been extracted from Dicke states and
their superpositions, the matrix elements are shown in
equation~(\ref{matrix elements}), our numerical results show that
$\min _{\theta,\phi}S_{A|B}\left( \theta,\phi\right)=S_{1}$. That
is, the minimum of $S_{A|B}\left( \theta,\phi\right)  $ over
$\theta$ is just obtained at the symmetric point
$\theta=\frac{\pi}{2}$ (see exemplifications in figures
\ref{Fig_Dicke_3D} and \ref{Fig_Ntheta_QD}). Therefore, we get the
compact expression of the PQD for the two-qubit reduced density
matrix (\ref{X structured}) as
\begin{equation}
\mathcal{D}(\rho^{AB})=S(\rho^{B})-S(\rho^{AB})+S_{1}, \label{QD X-state}%
\end{equation}
where $S(\rho^{AB})$, $S(\rho^{B})$ and $S_{1}$ are given in
equations\thinspace (\ref{S_AB (X)}), (\ref{S_B (X)}) and (\ref{S_0
and S_1}) respectively. In the following, we will apply
equation~(\ref{QD X-state}) to study the pairwise quantum
correlation of Dicke states and their superpositions.

Meanwhile, the PEoF for the reduced density matrix (\ref{X
structured}) is given by equation\thinspace(\ref{formation}) with
the concurrence expressed in~\cite{Vidal06},
\begin{equation}
C=\left\{
\begin{array}
[c]{cc}%
2(|u|-y),  |u|\geq y;\\
2(y-\sqrt{v_{+}v_{-}}),  y\geq\sqrt{v_{+}v_{-}}.
\end{array}
\right.  \label{Concurrence dicke}%
\end{equation}

\section{Symmetric multi-qubit states\label{sec:symmetric}}

The state (\ref{X structured}) can be obtained from the two-qubit
reduced density matrix of Dicke states or their superposition
states. In this case, the elements of the density matrix can be
expressed in terms of the expectation values of the collective spin
operators
\begin{equation}
\fl v_{\pm}=\frac{N^{2}-2N+4\left\langle J_{z}^{2}\right\rangle
\pm4\left\langle J_{z}\right\rangle (N-1)}{4N(N-1)},~~ y
=\frac{N^{2}-4\left\langle J_{z}^{2}\right\rangle }{4N(N-1)}
,~~u=\frac{\left\langle J_{+}^{2}\right\rangle }{N(N-1)}.
\label{matrix elements}
\end{equation}
where the collective spin operators are defined as $J_{\gamma}=\sum_{i=1}%
^{N}\frac{\sigma_{i\gamma}}{2}$ $\left(  \gamma=x,y,z\right)  $ with
$N$ the total spin number and $\sigma_{i\gamma}$ the pauli operator
on $i$-th site of spin. In the following, in terms of the PQD shown
in equation~(\ref{QD X-state}), we would like to study the pairwise
correlation of the X states (\ref{X structured}) with the special
elements shown in equation~(\ref{matrix elements}). In addition, the
results will be compared with those obtained from PEoF, which is
obtained by inserting equation~(\ref{Concurrence dicke}) into
equation~(\ref{formation}).

\subsection{Dicke state}\label{subsec-D}
A $N$-qubit Dicke state is defined as
\begin{equation}
\left\vert n\right\rangle _{N}=\left\vert \frac{N}{2},-\frac{N}{2}%
+n\right\rangle _{N},n=0,...,N, \label{Def1}
\end{equation}
and $\left\vert 0\right\rangle _{N}=\left\vert \frac{N}{2},-\frac{N}%
{2}\right\rangle $ indicates that all spins are pointing down. $N$
is the total spin number, and $n$ is the excitation number of spins
\cite{Dicke54}. The expressions for the relevant spin expectation
values can be easily obtained as~\cite{Wang02},
\begin{equation}
\langle J_{z}\rangle =n-\frac{N}{2},~~\langle J_{z}^{2}\rangle
=(n-\frac{N}{2})^{2},~~\langle J_{+}^{2}\rangle =0.
\end{equation}
From equation\thinspace (\ref{matrix elements}), it is easy to see
that the matrix elements of the reduced density matrix $\rho^{AB}$
are given by
\begin{center}
\begin{equation}
v_{+} =\frac{n(n-1)}{N(N-1)},~~v_{-}=\frac{(N-n)(N-n-1)}{N(N-1)},~~
y =\frac{n(N-n)}{N(N-1)},~~u=0. \label{Dicke_Me}%
\end{equation}
\end{center}

In figure\thinspace\ref{Fig_Dicke_3D}, we give a numerical check
that, for any $N$ and $n$, the minimum of the conditional entropy
$\min_{\phi}S_{A|B}(\theta,\phi)$ [as shown in equation~(\ref{QCE})]
are always obtained at $\theta=\pi/2$. The expression of PQD for
Dicke states can be obtained by inserting equation~(\ref{Dicke_Me})
into equation\thinspace(\ref{QD X-state}).

The corresponding expression of PEoF is obtained by inserting into
equation~(\ref{Concurrence dicke}) and equation~(\ref{formation}).
The two expressions are a little lengthy, so we don't explicitly
write them down.

\begin{figure}[ptb]
\begin{center}
\includegraphics[height=6cm,
width=10cm]{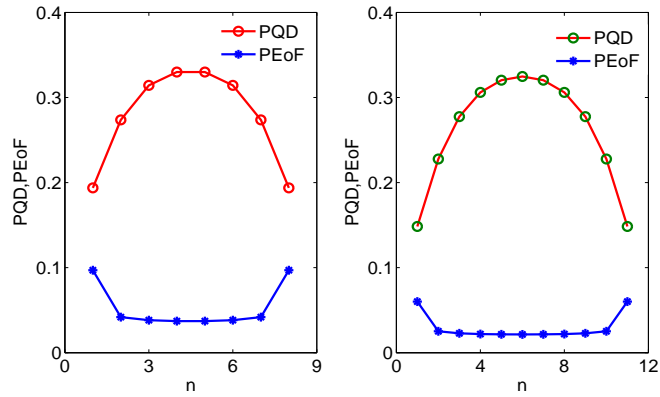}
\end{center}
\caption{(Color online) PQD and PEoF of Dicke state $|n\rangle_{N}$
for different excitation number $n\in\left[  1,N-1\right]  $ with
the fixed particle number $N=9$
(left) and $12$ (right).}%
\label{Fig_Dicke_QD_EOF_n}%
\end{figure}

First, we fixed the spin number $N$ to see the behavior of PQD and
PEoF for different excitation number $n$. For $n=0$ or $N$, the
Dicke state becomes a product state, which has zero PQD and zero
PEoF, i.e.,
\begin{equation}
\mathcal{D}(\rho_{\left\vert n\right\rangle
_{N}}^{AB})=E_{F}(\rho_{\left\vert n\right\rangle _{N}}^{AB})=0.
\end{equation}
In the following, we are interested in the cases that
$n\in\lbrack1,N-1]$. As shown in figure~\ref{Fig_Dicke_QD_EOF_n},
when $N$ is even, the PQD reaches to its maximum at $n=\frac{N}{2}$,
and the Dicke state reads $\left\vert n\right\rangle _{N}=\left\vert
\frac{N}{2}\right\rangle _{N}$, which has equal numbers of spins
pointing up and down. When $N$ is odd, the PQD arrives its maximum
at the point $n=\frac{N\pm1}{2}$, where the
Dicke state $\left\vert n\right\rangle _{N}=\left\vert \frac{N}{2}%
\pm\frac{1}{2}\right\rangle _{N}$ has the minimum different numbers
of spins pointing up and down. However, no matter $N$ is even or
odd, the maximum value of PEoF is obtained at the points $n=1$ or
$N-1$, where the Dicke states have the maximum different numbers of
spins pointing up and down, which are identical with the $W$ state
as shown in the equation\thinspace(\ref{G_W_state}). The different
behaviors of PQD and PEoF come from the different definitions of
them, as illustrated in Sec.~\ref{sec:qc}. Therefore, a state with
maximum quantum correlation may not have maximum quantum
entanglement.

\begin{figure}[ptb]
\begin{center}
\includegraphics[
height=6cm, width=10cm ]{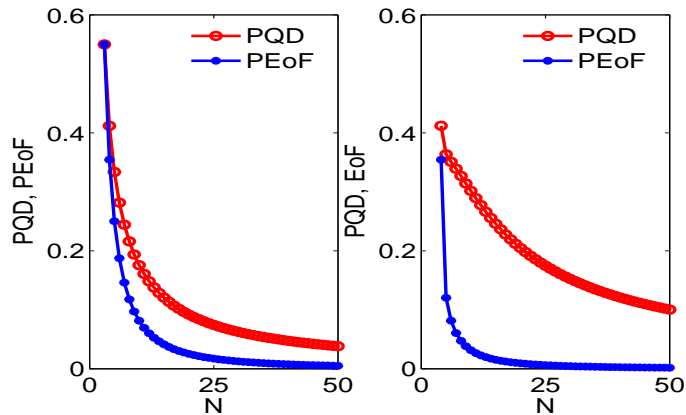}
\end{center}
\caption{(Color online) PQD and PEoF of Dicke state $|n\rangle_{N}$
for different particle
numbers $N\geq3$ with fixed excitation $n=1$ (left) and $3$ (right).}%
\label{Fig_Dicke_QD_EoF_N}%
\end{figure}

\begin{figure}[ptb]
\begin{center}
\includegraphics[
height=5cm, width=8cm ]{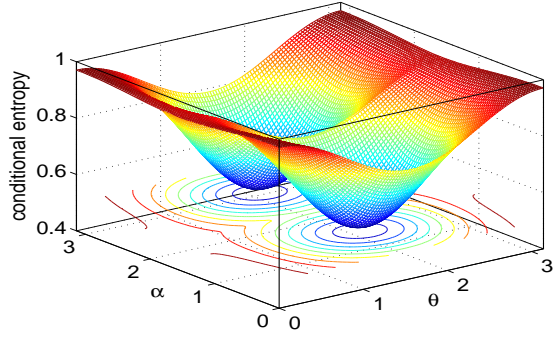}
\end{center}
\caption{(Color online) Conditional entropy
$\min_{\phi}S_{A|B}(\theta,\phi)$ [as shown in equation~(\ref{QCE})]
of the superposition of Dicke states $|\psi _{D}\rangle$ as a
function of the angles $\alpha$ and $\theta$ for fixed $N=50$ and
$n=30$. The minimum of $\min_{\phi}S_{A|B}(\theta,\phi)$ is obtained
at $\theta=\pi/2$ for fixed $n$, $N$ and $\alpha$.}
\label{Fig_Ntheta_QD}
\end{figure}

Second, we discuss the behaviors of PQD and PEoF for different spin
number $N$ with fixed excitation number $n$. For $n=1$, the Dicke
state becomes a generic $W$ state
\begin{equation}
\left\vert 1\right\rangle _{N}=\frac{1}{\sqrt{N}}(\left\vert
11\cdots 10\right\rangle +\left\vert 11\cdots
01\right\rangle+\cdots+\left\vert 01\cdots11\right\rangle ),
\label{G_W_state}%
\end{equation}
and the PQD reduces to
\begin{equation}
\mathcal{D}(\rho^{AB}_{\left\vert 1\right\rangle _{N}})=H\left(
\frac{1}{N}\right)
-H\left(  \frac{2}{N}\right)  +H\left(  \frac{N+\sqrt{(N-2)^{2}+4}}%
{2N}\right)  . \label{QDn=1}%
\end{equation}
On the other hand, the corresponding PEoF is given by
\begin{equation}
E_{F}(\rho^{AB}_{\left\vert 1\right\rangle _{N}})=\max_{n\in\lbrack1,N-1]}E_{F}(\rho_{\left\vert n\right\rangle _{N}}%
^{AB})=H\left(  \frac{N+\sqrt{N^{2}-4}}{2N}\right)  . \label{MAX_EoF}%
\end{equation}
It is easy to check that
\begin{equation}
\mathcal{D}(\rho^{AB}_{\left\vert 1\right\rangle _{N}})\geq
E_{F}(\rho^{AB}_{\left\vert 1\right\rangle _{N}}).
\end{equation}
For $n\geq1$, the similar relation between the PQD and the PEoF are
numerically shown in figure~\ref{Fig_Dicke_QD_EoF_N}. Both of them
decrease as the particle number $N$ increases, but the PQD reduces
more slowly than PEoF. That is, for an arbitrary Dicke state, the
general pairwise quantum correlation characterized by the PQD is
more robust against the increasing of the total particle number of
the Dicke ststes. This result is consistent with those obtained for
the reduced two-qubit states under a decoherence environment
\cite{Werlang09,Maziero10,Fanchini09}.

\subsection{Superposition of Dicke states}\label{subsec-SD}

Then we consider a simple superposition of Dicke states as%
\begin{equation}
\left\vert \psi_{D}\right\rangle =\cos\alpha\left\vert
n\right\rangle _{N}+e^{i\delta}\sin\alpha\left\vert n+2\right\rangle
_{N},
\label{Dicke_state}%
\end{equation}
where $n=0,...,N-2$, the angle $\alpha\in\lbrack0,\pi)$ and the
relative phase $\delta\in\lbrack0,2\pi)$. The expressions of the
relevant spin expectations
are%
\begin{eqnarray}
\left\langle J_{z}\right\rangle =\left( n-\frac{N}{2}\right) \cos
^{2}\alpha+\left( n+2-\frac{N}{2}\right)\sin^{2}\alpha, \nonumber\\
\left\langle J_{z}^{2}\right\rangle  =\left( n-\frac{N}{2}\right)
^{2}\cos^{2}\alpha+\left( n+2-\frac{N}{2}\right)^{2}\sin^{2}\alpha ,\nonumber\\
\left\langle J_{+}^{2}\right\rangle =\frac{1}{2}e^{i\delta}\sin
2\alpha\sqrt{\mu_{n}},\label{Expections}
\end{eqnarray}
with $\mu_{n}  =(n+1)(n+2)(N-n)(N-n-1)$.

For the state $\left\vert \psi_{D}\right\rangle $, the minimization
of the conditional entropy $S_{A|B}\left(  \theta,\phi\right)  $
over the measurement phase $\phi$ is obtained at
$\phi=\frac{1}{2}\delta$ or $\frac{1}{2}\delta +\pi$ according to
the equation\thinspace(\ref{min_phi}). Thus, the final expression of
the PQD is independent of the superposition phase $\delta$ from the
equation (\ref{k(theta)}). In figure 4, we also numerically check
that the conditional entropy $\min_{\phi}S_{A|B}(\theta,\phi)$ [as
shown in equation~(\ref{QCE})] of the state $\left\vert
\psi_{D}\right\rangle $ arrives its minimum at
$\theta=\frac{\pi}{2}$ for fixed $N$, $n$, and $\alpha$. Therefore,
the analytical expression\thinspace (\ref{QD X-state}) for the
superpositions of Dicke states is still valid.

There is a symmetry property of the PQD (or PEoF) for $\left\vert
\psi _{D}\right\rangle $. From equation
\thinspace(\ref{Expections}), we observe that if we let
$n\rightarrow N-n-2$ and $\alpha\rightarrow\frac{\pi}{2}+\alpha$,
then $\left\langle J_{z}\right\rangle \rightarrow-\left\langle
J_{z}\right\rangle $
and $\left\langle J_{+}^{2}\right\rangle \rightarrow-\left\langle J_{+}%
^{2}\right\rangle $, and furthermore $v_{+}\rightarrow v_{-}$ , $v_{-}%
\rightarrow v_{+}$ and $u\rightarrow-u$. Finally, two meaningful
relations are
obtained as%
\begin{equation}
\mathcal{D}(N,n,\alpha,\delta)=\mathcal{D}\left(  N,N-(n+2),\frac{\pi}%
{2}+\alpha,\phi\right)  , \label{Dicke_Symmetry_QD}%
\end{equation}
and
\begin{equation}
E_{F}(N,n,\alpha,\delta)=E_{F}\left(  N,N-(n+2),\frac{\pi}{2}+\alpha
,\phi\right)  . \label{Dicke_Symmetry_EoF}%
\end{equation}
That is, both the PQD and PEoF are symmetrical about $n=N/2-1$ and
$\alpha=-\pi/4$. These are useful properties for the following
analysis.

Next we would like to discuss the relation between the PQD and PEoF
for the state $\left\vert \psi_{D}\right\rangle $. First, when $N=2$
and $n=0$, the superposition of Dicke states becomes a two-qubit
GHZ-like state,
\begin{equation}
\left\vert \psi_{D}\right\rangle =\cos\alpha\left\vert
2\right\rangle +e^{i\delta}\sin\alpha\left\vert 0\right\rangle .
\end{equation}
This is a bipartite pure state so that the PQD equals PEoF, i.e.,
\begin{equation}
\mathcal{D}=E_{F}=H\left(  \cos^{2}\alpha\right)  .
\end{equation}
This is also the result for the case that $N=2$ and $n=1$ according
to the symmetry properties \thinspace(\ref{Dicke_Symmetry_QD})
and\thinspace (\ref{Dicke_Symmetry_EoF}). In fact, for a general
multi-qubit GHZ-like state with $N\geq{3}$
\begin{equation}
|\rm{GHZ}\rangle=\cos\alpha\left\vert N\right\rangle _{N}+e^{i\delta}%
\sin\alpha\left\vert 0 \right\rangle _{N}, \label{GGHZ}%
\end{equation}
there is no quantum correlation for the two-qubit reduced state,
i.e.,
\begin{equation}
\mathcal{D}=E_{F}=0.
\end{equation}
In summary, for a multi-qubit GHZ-like state, the PQD is always equal
to the PEoF. Their values are always zero except for $N=2$.

Second, when $N=3$, it is interesting that, for any $n$, the state
$\left\vert \psi_{D}\right\rangle $ has equal PQD and PEoF. In fact,
for $N=3$ and
$n=0$, the elements of the reduced density matrix are%
\begin{equation}
v_{+}=y=\frac{\sin^{2}\alpha}{3},v_{-}=\cos^{2}\alpha,u=\frac{\sqrt
{3}e^{i\delta}\sin2\alpha}{6}.
\end{equation}
And for $N=3$ and $n=1$, they are%
\begin{equation}
v_{-}=y=\frac{\sin^{2}\alpha}{3},v_{+}=\cos^{2}\alpha,u=\frac{\sqrt
{3}e^{i\delta}\sin2\alpha}{6}.
\end{equation}
Both of them satisfy the relation that
\begin{equation}
v_{+}=y,~~~~ v_{+}v_{-}=|u|^{2}. \label{relation_N=3}%
\end{equation}
This ensures the joint entropy $S\left(
\rho_{|\psi\rangle_{D}}^{AB}\right)$ equals the reduced entropy
$S\left( \rho_{|\psi\rangle _{D}}^{A}\right)$, i.e.,
\begin{equation}
S\left(  \rho_{|\psi\rangle_{D}}^{AB}\right)  =S\left(
\rho_{|\psi\rangle _{D}}^{A}\right)  .
\end{equation}
Therefore, the PQD is
\begin{equation}
\mathcal{D}=\min_{\theta,\phi}S_{A|B}(\theta,\phi)=S_{1}=H\left(
\frac{1+\kappa_{1}}{2}\right)  ,
\end{equation}
with%
\begin{equation}
\kappa_{1}=\sqrt{(v_{+}-v_{-})^{2}+4(y+|u|)^{2}}.
\end{equation}
Meanwhile, from equation \thinspace(\ref{relation_N=3}), the
corresponding concurrence is given by
\begin{equation}
C^{2}=4(y-|u|)^{2}. \label{Concurrence}%
\end{equation}
Obviously,
\begin{equation}
\kappa_{1}=\sqrt{1-C^{2}}.
\end{equation}
Thus, PQD is equal to PEoF, i.e.,
\begin{equation}
\mathcal{D}=E_{F}.
\end{equation}
Combining the symmetry properties\thinspace(\ref{Dicke_Symmetry_QD})
and\thinspace(\ref{Dicke_Symmetry_EoF}), this equivalence can be
extended to any $n$ for $N=3$. This indicates that for certain kinds
of mixed state, quantum entanglement may also represents the whole
quantum correlation.

\begin{figure}[ptb]
\begin{center}
\includegraphics[
height=6cm, width=8cm ]{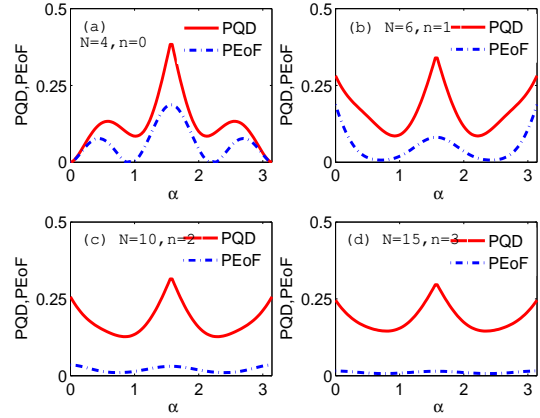}
\end{center}
\caption{(Color online) PQD and PEoF for superposition of Dicke
state $|\psi_{D}\rangle$\ as
the function of $\alpha$ for different $N\geq{4}$ and $n$. }%
\label{Fig_super_Dicke_N_n}%
\end{figure}

Both of the above two cases with $N=2$ and $N=3$ show that PQDs are
equal to PEoFs for pure and some special mixed two-qubit states.
However, when $N\geq4$, the equivalence does not always hold, as
shown in figure \thinspace \ref{Fig_super_Dicke_N_n}. It seems that
the PQD is always greater than or equal to PEoF. This is similar to
the case of Dicke states. In addition, the results are a little more
complex than that of Dicke state. Every possible cases exists. The
state with maximum (minimum) PEoF may have (or not have) the maximum
(minimum) PQD as shown in figure \thinspace\ref{Fig_super_Dicke_N_n}
(a) and (b) for small particle number $N$. While for large $N$, as
shown in figure\thinspace\ref{Fig_super_Dicke_N_n} (c) and (d), the
state with maximum (minimum) PEoF approaches to the one with the
maximum (minimum) PQD.

\subsection{Spin coherent states}

Finally, we consider two more complex superpositions of Dicke
states, i.e., the superpositions of the form
$\sum_{n}C_{n}\left\vert n\right\rangle _{N}$ with the excitation
numbers $n$ even and odd respectively. They can be given by
SCSs\thinspace\cite{Radcliffe71,Gerry98,Arecchi72}
\begin{equation}
\left\vert \eta\right\rangle _{\pm}=\frac{1}{\sqrt{2(1\pm\gamma^{N})}%
}(\left\vert \eta\right\rangle \pm\left\vert -\eta\right\rangle ),
\label{even and odd}%
\end{equation}
with $\gamma=(1-\eta^{2})/(1+\eta^{2})$. The $``\pm``$ corresponds to
the so called even SCSs (ESCSs) and odd SCSs (OSCSs), respectively.
A SCS is obtained by a rotation of the Dicke state $\left\vert
0\right\rangle _{N},$
\begin{equation}
\left\vert
\eta\right\rangle=(1+\eta^{2})^{-N/2}\exp{(J_{+}\eta)}|0\rangle_{N}
=(1+\eta^{2})^{-N/2}\sum_{n=0}^{N}\left(
\begin{array}
[c]{c}%
N\\
n
\end{array}
\right)  ^{1/2}\eta^{n}\left\vert n\right\rangle _{N},
\end{equation}
where the parameter $\eta\in\lbrack0,1]$. The expectations for the
ESCSs and OSCSs are
\begin{eqnarray}
\left\langle J_{z}\right\rangle _{\pm} =\frac{-N}{2}\frac{\gamma\pm
\gamma^{N-1}}{1\pm\gamma^{N}},\nonumber\\
\left\langle J_{z}^{2}\right\rangle _{\pm}  =\frac{N^{2}}{4}\pm
\frac{N(N-1)\eta^{2}\upsilon_{\eta}^{\mp}}{1\pm\gamma^{N}},\nonumber\\
\left\langle J_{+}^{2}\right\rangle _{\pm}  =\pm\frac{N(N-1)\eta
^{2}\upsilon_{\eta}^{\pm}}{1\pm\gamma^{N}}, \label{SCS expection}%
\end{eqnarray}
with
$\upsilon_{\eta}^{\pm}=\gamma^{N}(1-\eta^{2})^{-2}\pm(1+\eta^{2})^{-2}$.

\begin{figure}[ptb]
\begin{center}
\includegraphics[
height=6cm, width=8cm ]{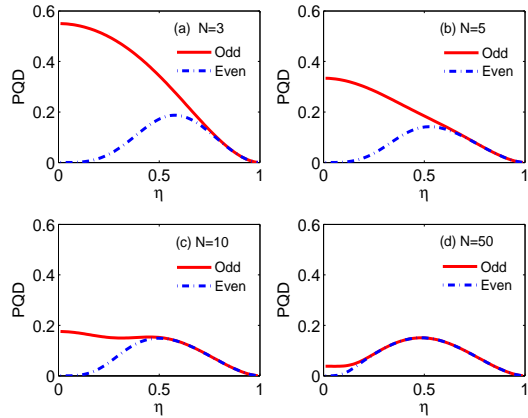}
\end{center}
\caption{(Color online) PQD for odd and even SCSs along with the
parameter $\eta\in\left[
0,1\right]  $ for different spin numbers $N$. }%
\label{Fig_SCSs_N}%
\end{figure}

Similar to the above two cases shown in Secs.~\ref{subsec-D} and
\ref{subsec-SD}, we also numerically check that the analytical
expression\thinspace(\ref{QD X-state}) of PQD can be reliably
accepted. Here the expression of PQD is so lengthy that we only
numerically show its behaviors for different parameters, as
displayed in figure\thinspace\ref{Fig_SCSs_N}. We see that, when
$\eta$ is small, the QD of OSCSs is always greater than that of
ESCSs for fixed particle number $N\geq3$. However, when $\eta$
becomes large and approaches $1$, they become gradually equal to
each other. Especially, when $N\gg1$, they coincide with each other
even for small $\eta$. All these results can be explained
analytically from the three special cases in the following.

When $\eta\rightarrow0$, the PQD of the OSCS is different from that
of the ESCS . The OSCS reduces to the Dicke state\thinspace
$\left\vert \eta=0\right\rangle _{-}=\left\vert 1\right\rangle _{N}$
(the $W$ state shown in equation\thinspace(\ref{G_W_state})), whose
PQD is nonzero as shown in equation\thinspace(\ref{QDn=1}).
Meanwhile, the ESCS reduces to product state $\left\vert
\eta=0\right\rangle _{+}=\left\vert 0\right\rangle _{N}$, then
$\mathcal{D}=0$, as shown in figure\thinspace\ref{Fig_SCSs_N}.

However, when\thinspace$\eta\rightarrow1$, the PQDs of OSCSs and
ESCSs coincide with each other. In this case, both of the OSCSs and
ESCSs reduces to GHZ states in the $x$
direction\thinspace\cite{Arecchi72},
\begin{equation}
\left\vert \eta=1\right\rangle _{\pm}=\frac{1}{\sqrt{2}}((\left\vert
N\right\rangle _{N})_{x}\pm(\left\vert 0\right\rangle _{N})_{x}).
\end{equation}
If $N\geq3$, the reduced density matrix of the OSCSs and ESCSs are
diagonalized
\begin{equation}
\rho^{AB}|_{\eta=1}=\frac{1}{2}\left\vert 00\right\rangle
\left\langle 00\right\vert +\frac{1}{2}\left\vert 11\right\rangle
\left\langle 11\right\vert ,
\end{equation}
and we have $\mathcal{D}=0$.

Finally, when $N\gg1$, for both the OSCSs and ESCSs, the reduced
density matrices are
almost independent of $N$,%
\begin{equation}
\rho^{AB}|_{N\gg1}=\frac{1}{(1+\eta^{2})^{2}}\left(
\begin{array}
[c]{cccc}%
 1& 0 & 0 & \eta^{2}\\
0 & \eta^{2} & \eta^{2} & 0\\
0 & \eta^{2} & \eta^{2} & 0\\
\eta^{2} & 0 & 0 & \eta^{4}%
\end{array}
\right)  ,
\end{equation}
and PQD can be given by%
\begin{equation}
\fl\mathcal{D}(\rho^{AB}|_{N\gg1}) =H\left(
\frac{1}{1+\eta^{2}}\right) -H\left(
\frac{2\eta^{2}}{(1+\eta^{2})^{2}}\right)  \nonumber +H\left(
\frac{(1+\eta^{2})^{2}+\sqrt{1+14\eta^{4}+\eta
^{8}}}{(1+\eta^{2})^{2}}\right)  .
\end{equation}
Then PQD will also become independent of $N$. As a result, PQDs of
OSCSs and ESCSs become close to each other even for small $\eta$, as
shown in figure \thinspace \ref{Fig_SCSs_N} (d). In particular, we
obtain $E_F(\rho^{AB}|_{N\gg1})=0$. This implies that QD is a more
general measure of quantum correlation than quantum entanglement.
For a set of quantum separable states, they do have quantum
correlations.

\begin{figure}[ptb]
\begin{center}
\includegraphics[
height=5cm, width=7cm]{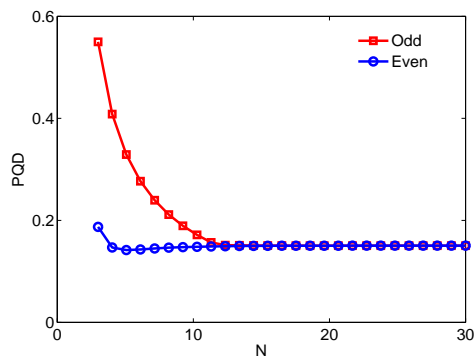}
\end{center}
\caption{(Color online) Maximum PQD of the ESCSs and OSCSs over the
parameter scale $\eta
\in\left[0,1\right]  $ for different $N\geq3$. }%
\label{Fig_SCSs_maximum}%
\end{figure}

In figure~\ref{Fig_SCSs_maximum}, we display the relation of maximum
PQDs between the OSCSs and ESCSs over the parameter scale
$\eta\in\left[ 0,1\right]  $. It shows that the maximum PQD of OSCSs
is always greater than or equal to that of ESCSs for fixed particle
number $N$. When $N\gg1$, their maximum values attain a constant,
which can also be seen from figure \thinspace\ref{Fig_SCSs_N}.

\section{\textbf{Conclusion}}

In terms of PQD, we have investigated the pairwise quantum
correlations in Dicke states and their superpositions in terms of
its two-qubit density matrix, whose elements may be complex. A
general expression for PQD is derived according to the numerical
proof. For the Dicke states, our analytical and numerical results
show that the PQD is always greater than or equal to the PEoF. This
further proves that QD is a more general measure of quantum
correlation than quantum entanglement. For the superpositions of
Dicke states, it is interesting that the QD is always equal to EoF
when $N=2$ (GHZ-like states) and $N=3$ (W states), which indicates
that for some kinds of mixed states, quantum correlation can also be
fully described by quantum entanglement. For OSCSs and ESCSs, the
former always show more quantum correlations than the latter. In
addition, the PQD is generally more robust against the enlarging of
the particle number of the multi-qubit states, which implies that PQD
has an advantage over pairwise entanglement in characterizing the
quantum correlation in the multi-qubit states. To some extent, the
pairwise quantum correlations of the reduced states may reflect the
multi-qubit quantum correlations of the global states.

\section*{Acknowledgments}

We are grateful to Xiao-Ming Lu, Xiaoqian Wang, and Jian Ma for
fruitful discussions. This work is supported by NSFC with grant
No.10874151, 10935010, 11025527, 60873119, NFRPC with grant No.
2006CB921205; the Fundamental Research Funds for the Central
Universities, and as well as the Superior Dissertation Foundation of
Shannxi Normal University (S2009YB03), and the Higher School
Doctoral Subject Foundation of Ministry of Education of China under
grant No. 200807180005.

\noappendix

\end{document}